# Thermo-hydro-mechanical modeling of an Enhanced geothermal system in a fractured reservoir using $CO_2$ as heat transmission fluid- A sensitivity investigation

Saeed Mahmoodpour[1], Mrityunjay Singh[1], Kristian Bär[1], Ingo Sass[1,2]

[1] Technische Universität Darmstadt, Institute of Applied Geosciences, Group of Geothermal Science and Technology, Schnittspahnstrasse 9, 64287 Darmstadt, Germany

[2] Darmstadt Graduate School of Excellence Energy Science and Engineering, Otto-Berndt-Strasse 3, 64287 Darmstadt, Germany

*Correspondence to*: **Saeed Mahmoodpour (saeed.mahmoodpour@tu-darmstadt.de)**

**Abstract**

Geothermal energy has the potential to support direct heat usage and electricity generation at low carbon footprint. Using $CO_2$ as heat transfer fluid can allow us to achieve negative carbon energy solution. In this study, geothermal energy extraction potential from a discretely fractured reservoir using $CO_2$ is assessed. Geothermal energy extraction process is a coupled thermo-hydro-mechanical (THM) mechanism and the geomechanical stresses involves thermoelasticity and poroelasticity. This study demonstrates a fully coupled THM mechanism for enhanced geothermal system (EGS) operations. A large number of parameters are involved in the THM mechanism and therefore, it becomes difficult to assess the key operating parameters to have better operating efficiency. We identified 22 input parameters that controls the THM mechanism. Therefore, under the Horizon 2020 project: Multidisciplinary and multi-contact demonstration of EGS exploration and Exploitation Techniques and potentials (H2020 MEET) we have performed sensitivity analysis to investigate the relative importance of these parameters that concentrates on three key objective parameters: thermal breakthrough time, mass flux and overall energy recovery. The important parameters controlling these three objective parameters are matrix permeability and fracture aperture whereas wellbore radius has an impact on mass flux and total energy recovery.

# 1. Introduction

Carbon dioxide release in Earth's atmosphere has detrimental impacts on climate change and global warming (IPCC 2005). The primary source of carbon dioxide generation is burning hydrocarbon fuels from stationary and mobile power generation sources. Sequestration of $CO_2$ in deep saline reservoirs is a viable solution to the anthropogenic emission where dissolution and capillary trapping acts in a short time span (Mahmoodpour et al. 2019; Singh et al. 2020a, 2021). However, $CO_2$ sequestration is a cost intensive approach and therefore, to limit this anthropogenic emission of $CO_2$ in the atmosphere, alternate energy sources are getting prime attention. Geothermal energy extraction is one such technology that can supply low carbon electrical and heating power. Circulation of cold water in deep and hot fractured earth curst produces hot water at the production outlet and thus geothermal energy is harnessed. Fresh water is highly reactive and a scarce resource in many regions across the world. Further, using $CO_2$ instead of water as heat carrying fluid may result in carbon negative geothermal energy source. Circulating fluid loss has a critical impact on efficiency of power generation. However, loss of $CO_2$ from the fractured reservoir may result in carbon geosequestration. $CO_2$ is a less reactive chemical compound with respect to water in the subsurface environment and the unwanted $CO_2$ emission from hydrocarbon based power plants can supply abundant amount of $CO_2$. Therefore, using $CO_2$ as heat tranfer fluid has two major advantages: offering a less reactive heat transfer fluid and $CO_2$ geological sequestration (Singh et al. 2020b). In this paper, we have developed a numerical model depicting geothermal energy extraction from a fractured reservoir using $CO_2$ as heat transfer fluid. A thermo-hydro-mechanical (THM) model is developed in this study. A comprehensive sensitivity anslysis is performed using Plackett-Burman methodology to understand the relative importance of parameters governing the THM process and identify the key parameters which has highest impact on temperature drop, mass flux and the total energy recovery measured at the production well.

For the first time, Brown (2000) proposed the use of supercritical $CO_2$ at Fenton Hill in New Mexico at 4 km depth and 260 °C temperature for enhanced geothermal operation. He proposed three major advantages of using $CO_2$ over water as heat transfer fluid: high density difference for working fluid between injection and production well due to higher thermal expansion coefficient, less chemical reactivity resulting in reduced scaling and corrosion in the piping, heat exchangers and possibility of EGS implementation in rocks with temperature greater than 374 °C. He also idealized a closed circulating loop of flowing $CO_2$ with production wellhead temperature as 250 °C. Further, Liu et al (2003) investigated the geochemical interaction of $CO_2$ during EGS operating conditions in granite and sandstone for temperature between 100 °C and 350 °C and found that dissolution increases with increase in $CO_2$ content. Additionally, they investigated the potential of $CO_2$ sequestration during EGS operation. Using numerical models, Pruess (2006) showed that with increase in injection well downhole temperature, heat mining rate increases and a pressure gradient reduction is possible with depth. This results in buoyant pressure drop which is necessary to push $CO_2$ towards the production well. $CO_2$ possess higher compressibility and expansivity compared to water. Due to this, $CO_2$ circulation in the rock consumes less power when compared to water. Further, viscosity of $CO_2$ is less than water, resulting in comparatively higher velocity for a given pressure gradient. A large fluctuation in the $CO_2$ mobility due to temperature and pressure compared to water governs unusual mass flux at the production well (Pruess 2008). He also proposed that placing a production well along the reservoir top will assist in delaying the gravity driven $CO_2$ plume arrival at the production well to avoid thermal breakthrough. However, this feature is noticeably higher in case of $CO_2$ and not in case of water. Later, Zhang et al. (2014) commended to use $CO_2$ instead of water due to higher thermal expansivity, and strong temperature and pressure dependent mobility of $CO_2$.

The circulating mass flux in the reservoir has a major role in quantification of produced heat at the production well. Luo et al. (2014) performed a comparative study between water and $CO_2$ based doublet for EGS and found that $CO_2$ requires double mass flux compared to water for the same heat production and at the same time pressure drop across the wells is small in case of $CO_2$. Later Liu et al. (2017) found that low flux between injection-production well produces high temperature $CO_2$ compared to high mass flux. Using numerical models, Bai et al. (2018) showed that increasing fracture roughness or the flux results in better heat transfer when $CO_2$ works as heat transferring fluid. Using a pipe-network based discrete fracture networks (DFNs) in a two-phase fluid system, Chen et al. (2019) showed that for a hydro-thermal process, increasing the $CO_2$ saturation in the reservoir before heat extraction could enhance the heat production performance. Tang et al. (2014) showed that large fracture size in a reservoir leads to higher energy production. Concurrently, small fracture aperture and small fluid velocity delays the thermal breakthrough time and therefore, it assists in maintaining higher production well temperature. On the other hand greater energy generation requires higher velocity and large fracture aperture.

When a fractured reservoir model is considered in such a way that fractures and the porous medium are considered as a single medium, a single temperature is sufficient to understand the energy balance. Here, the fractured reservoir reaches the porous medium temperature instantaneously (Yang et al. 2009). Therefore, the fractures and the rock matrix are in thermodynamic equilibrium at all times in this model. This approach is known as local thermal equilibrium (LTE). Shaik et al. (2011) found that LTE models underestimate the fluid production rate. When considering the local thermal non-equilibrium (LTNE) model, the energy production depends on volumetric heat transfer coefficient. Here, the volumetric heat transfer coefficient is the product of specific surface area and the internal heat transfer coefficient (Chen et al. 2019; Wang et al. 2019; Zhang et al. 2014). Later Zhang et al. (2014) observed that the thermal breakthrough time and produced fluid temperature prediction using LTNE models are considerably different than LTE models.

Injection of cold $CO_2$ in a fractured reservoir increases the formation pressure leading to geomechanical deformation. Shao et al. (2014) found that reservoir cooling develops multiple micro-structural alteration in the grain size and shape. From the existing literature on water based enhanced geothermal systems, Zhang and Zhao (2019) reported that faults remain intact if the contact strength is below the shear stress. Sun et al. (2019) developed a THM model for EGS using $CO_2$ as working fluid in COMSOL Multiphysics and proposed a cyclic alteration of pressure at the injection and production wells. They suggested that pressure alteration at the wells improves heat extraction efficiency. Guo et al. (2019) developed a THM model for $CO_2$-EGS consisting fractures with different length scales and demonstrated that fracture connectivity has to play a major role in heat extraction performance. Later Liao et al. (2020) performed field scale numerical modeling and injected supercritical $CO_2$ for fracture network generation in a geothermal reservoir. They compared the impact of water and $CO_2$ usage as working fluid and found that both fluids indicate a positive correlation for injection rate with respect to injection pressure and thermal capacity whereas a negative correlation was observed for the produced temperature with respect to injection rate.

Enhanced geothermal system operations are governed by a large number of parameters. Fracture characteristics, thermophysical fluid properties, rock matrix conditions and the geological loading conditions are simultaneously responsible for EGS operations. Therefore, understanding the impact of these parameters on three objectives: thermal breakthrough at the production well, mass flux output and the total energy recovery is critical before designing and operating an EGS site. However, literature on the sensitivity of parameters affecting these three objectives is limited. In this paper, first we developed a THM model for EGS operation using $CO_2$ as working fluid. Next, we considered 22 indepnedent parameters that controls the fracture, rock matrix and $CO_2$ properties along with initial and boundary conditions and designed a statistical approach to understand the sensitivity of these 22 parameters. The present study proposes a novel sensitivity analysis that identifies key parameters controlling three main objectives of the EGS operations when $CO_2$ is the working fluid. Lastly, we compared the results obtained from sensitivity analysis for $CO_2$ and water (Mahmoodpour et al. 2021).

# 2. Methods

**2.1 Model schematic, initial and boundary conditions**

A discretely fractured reservoir is considered based on the outcrop fractures mapped from Otsego County in New York State for THM study in this work (Mahmoodpour and Masihi 2016). Figure 1 shows a two dimensional planar geometry of size 1000 m × 600 m with 440 fractures. We assumed all fractures as interior boundaries to eliminate the requirements of high aspect ratio geometry. All side boundaries are displacement contrained in the normal directions to its sides. Furthermore, no heat and mass transfer is assumed across the boundaries.

## 2.2 Governing equations

The mass conservation equation to couple pore volume variation and fluid temperature for a porous medium can be written as:

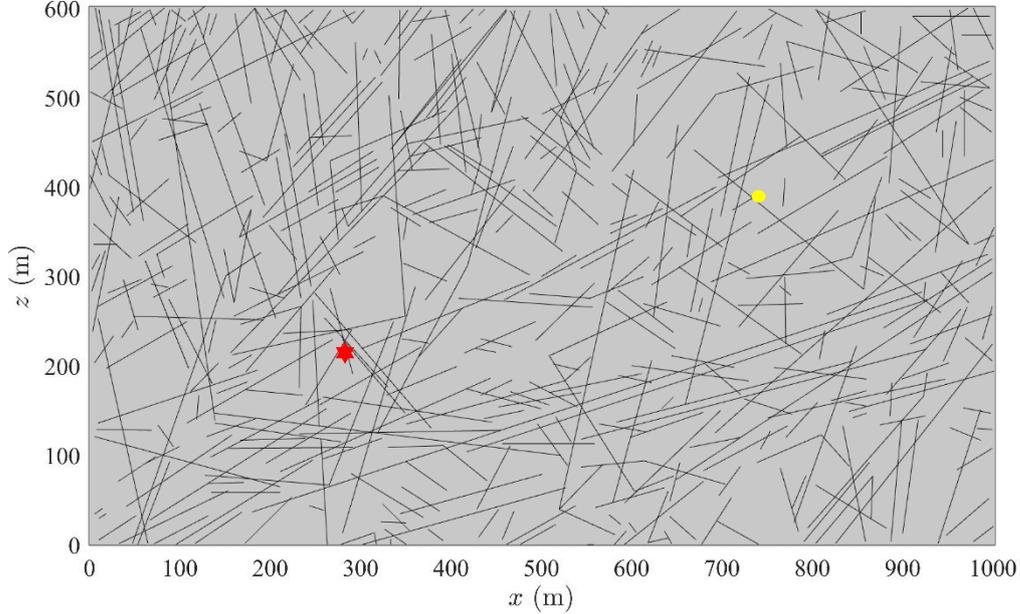

*Figure 1. Geometry of the model. Here, yellow circle is the injection well and red star is production well.*

$$\rho_1(\phi_m S_1 + (1-\phi_m)S_m)\frac{\partial p}{\partial t} - \rho_1(\alpha_m(\phi_m \beta_1 + (1-\phi_m)\beta_m))\frac{\partial T}{\partial t} + \rho_1 \alpha_m \frac{\partial \varepsilon_V}{\partial t} = \nabla \cdot \left(\frac{\rho_1 k_m}{\mu}\nabla p\right) \quad [1]$$

We have ignored the fluid flow along the fracture width as the dimension of fracture length is in the order of m and the fracture aperture is in the order of mm. The flow along the internal fractures can be written using the tangential derivatives as follow:

$$\rho_1(\phi_f S_1 + (1-\phi_f)S_{mf})e_h \frac{\partial p}{\partial t} - \rho_1(\alpha_f(\phi_f \beta_1 + (1-\phi_f)\beta_f))e_h \frac{\partial T}{\partial t} + \rho_1 \alpha_f e_h \frac{\partial \varepsilon_V}{\partial t} =$$
$$\nabla_T \cdot \left(\frac{e_h \rho_1 k_f}{\mu}\nabla_T p\right) + n \cdot Q_m \quad [2]$$

We assumed that the diameter of the wells are small and therefore they are represented by lines. Furthermore, we adopted local thermal non-equilibrium (LTNE) approach for the heat exchange between fluid and the rock matrix. Heat conduction dominates heat exchange in both the rock matrix as well as in the fracture. The governing equation for heat exchange in rock matrix is defined by:

$$(1-\phi_m)\rho_m C_{p,m}\frac{\partial T_m}{\partial t} = \nabla \cdot \left((1-\phi_m)\lambda_m \nabla T_m\right) + q_{ml}(T_l - T_m) \quad [3]$$

The governing equation for heat exchange in the fracture can be written as:

$$(1-\phi_f)e_h \rho_f C_{p,f}\frac{\partial T_m}{\partial t} = \nabla_T \cdot \left((1-\phi_f)e_h \lambda_f \nabla_T T_m\right) + e_h q_{fl}(T_l - T_m) + n \cdot (-(1-\phi_m)\lambda_m \nabla T_m) \quad [4]$$

The convective heat transfer equation for the pore fluid can be expressed by the following energy balance equation:

$$\phi_m \rho_l C_{p,l}\frac{\partial T_l}{\partial t} + \phi_m \rho_l C_{p,l}\left(-\frac{k_m \nabla p}{\mu}\right) \cdot \nabla T_l = \nabla \cdot (\phi_m \lambda_l \nabla T_l) + q_{ml}(T_m - T_l) \quad [5]$$

The coupled equation for heat exchange between the rock matrix and the fracture can be written as:

$$\phi_f e_h \rho_l C_{p,l} \frac{\partial T_l}{\partial t} + \phi_f e_h \rho_l C_{p,l}(-\frac{k_f \nabla_T p}{\mu}).\nabla_T T_l = \nabla_T.(\phi_f e_h \lambda_l \nabla_T T_l) + e_h q_{fl}(T_m - T_l) + n.(-\phi_l \lambda_l \nabla T_l)$$
[6]

where $u_f = -\frac{k_f \nabla_T p}{\mu}$ is the mass flux in fractures; the heat flux $n.q_l = n.(-\phi_l \lambda_l \nabla T_l)$ indicates heat exchange of the fluid between rock matrix and the fracture.

The thermodynamic properties of $CO_2$ is incorporated in Eq. [1-6] are temperature dependent as listed in Eq. [10-14].

### 2.2.1 Stress

We have considered linear elasticity model to investigate THM process proposed in this study. We can write the effective stress, and thermal stress theory, and the thermo-poroelastic equation describing the stress-strain relationship of porous media as follows:

$$\sigma_{ij} = 2G\varepsilon_{ij} + \lambda tr\varepsilon \delta_{ij} - \alpha_p p \delta_{ij} - K'\beta_T T \delta_{ij} \qquad [7]$$

In the above equation, $\beta_T = \phi_l \beta_l + (1-\phi_m)\beta_m$ is the volumetric thermal expansion coefficient of porous media and $\sigma_{eff}^{ij} = \sigma_{ij} + \alpha_p p \delta_{ij}$ is the effective stress.

Finally, we can write the porous medium deformation equation as follow:

$$Gu_{i,jj} + (G+\lambda)u_{j,ji} - \alpha_p p_{,i} - K'\beta_T T_{,i} + f_i = 0 \qquad [8]$$

We have assumed Barton and Bandis model to describe the opening and closure processes of the stress-dependent fracture (Bandis et al. 1983; Barton et al. 1985). The initial fracture aperture variation under in-situ stresses can be written as:

$$\Delta e_n = \frac{e_0}{1 + 9\frac{\sigma_{eff}^n}{\sigma_{nref}}} \qquad [9]$$

### 2.2.2 Fluid properties

The thermophysical properties of $CO_2$ such as dynamic viscosity, specific heat capacity, density, and thermal conductivity as a function of temperature are written below:

Dynamic viscosity

$$\mu = -1.48502597 \times 10^{-6} - 6.46786388 \times 10^{-8} \times T^1 - 3.66170537 \times 10^{-11} \times T^2 + 1.24501217 \times 10^{-14} \times T^3 \qquad (220 - 600\ K) \qquad [10]$$

Specific heat capacity

$$C_p = 459.913258 + 1.86487996 \times T^1 - 0.00212921519 \times T^2 + 1.22488004 \times 10^{-6} \times T^3$$

$$(220 - 600\ K) \qquad [10]$$

Density which is a function of pressure and temperature can be written as:

$$\rho = pA \times 0.04401/RT \qquad [13]$$

In the above equation, $pA$ is the absolute pressure and $R$ is the molar gas constant. Thermal conductivity for $CO_2$ can be written as,

$$\kappa = -0.00132472616 + 4.13956923 \times 10^{-5} \times T^1 + 6.70889081 \times 10^{-8} \times T^2 - 2.11083153 \times 10^{-11} \times T^3 \qquad [14]$$

COMSOL Multiphysics version 5.5 (COMSOL) is used to perform numerical modeling of THM processes. It uses a finte element method to solve general purpose partial differential equations. The complete mesh contains 112, 818 domain elements and 13,071 boundary elements. For the numerical modeling purpose, we have used an scaled absolute tolerance of magnitude $10^{-8}$ and automatic time step constraint. We assumed Backward Differentiation Formula (BDF) for timestepping with maximum BDF order as 2 and minimum BDF order as 1. Further, we have validated our model with a soil thermal consolidation model as demonstrated by Bai (2005) in Mahmoodpour et al. (2021).

## 3. Results and Discussions

Results from fully coupled THM simulation for geothermal energy extraction from a fractured geothermal reservoir are presented in this section. First, we have demonstrated the coupled thermo-poroelastic stress impact on fluid flow and heat exchange between the rock matrix and the fractures. Next, we have presented the Plackett-Burman method for the sensitivity analysis based on 48 reservoir simulations and identified critical parameters influencing three key objectives: thermal breakthrough time, mass flux, and overall energy recovery.

### 3.1. THM modeling

In this section, we considered a discretely fractured reservoir for numerical simulation of the geothermal energy extraction potential by injecting cold $CO_2$ and producing hot $CO_2$. The input parameters for the representative case are listed in Table 2.

*Table 2: Parameters used to investigate the long-term EGS potential fromfractured geothermal reservoir. The minimum and maximum values for parameters listed in Table 3. The minimal value is represented using -1 whereas maximum value is denoted by 1.*

| Parameter | E | ν | ρr | S1 | S2 | Pi | Pj | φr | kr | φf | f | Ap | σs | wr | λr | λf | Cr | Cf | Ti | α | β | Tj |
|---|---|---|---|---|---|---|---|---|---|---|---|---|---|---|---|---|---|---|---|---|---|---|
| value | 1 | -1 | 1 | -1 | 1 | -1 | -1 | 1 | 1 | 1 | 1 | -1 | 1 | 1 | 1 | 1 | 1 | -1 | -1 | -1 | -1 | 1 |

*Table 3: Parameters used to investigate the sensitivity study. The minimal value is represented using -1 whereas maximum value is denoted by 1.*

| Symbol for Numerical model | Symbols for Sensitivity study | Term | Parameter | Range |
|---|---|---|---|---|
| E | A | E | Young's modulus | 20 GPa – 40 GPa |
| ν | B | PR | Poisson's ratio | 0.2 – 0.3 |
| ρr | C | RD | Rock density | $2500 \frac{kg}{m^3}$ - $2800 \frac{kg}{m^3}$ |
| S1 | D | S1 | Horizontal stress | 30 MPa – 50 MPa |
| S2 | E | S2 | Vertical stress | 30 MPa – 50 MPa |
| pi | F | IP | Initial pressure | 20 MPa – 30 MPa |
| pj | G | InjP | Injection pressure | 50 MPa – 60 MPa |
| φr | H | RP | Rock porosity | 0.05 – 0.2 |
| kr | J | Rper | Rock permeability | 2 mD – 5 mD |
| φf | K | FP | Fracture porosity | 0.3 – 0.5 |
| f | L | FR | Fracture roughness | 1 – 2 |
| Ap | M | FA | Fracture aperture | 0.1 mm – 0.2 mm |
| σs | N | ES | Closure stress | 100 MPa – 150 MPa |
| wr | O | WR | Wellbore radius | 0.1 m – 0.2 m |

| λr | P | RT | Rock thermal conductivity | $2 \frac{W}{m \times K} - 3 \frac{W}{m \times K}$ |
|---|---|---|---|---|
| λf | Q | FT | Fracture thermal conductivity | $1.5 \frac{W}{m \times K} - 2 \frac{W}{m \times K}$ |
| Cr | R | RSH | Rock specific heat capacity | $800 \frac{J}{kg \times K} - 1000 \frac{J}{kg \times K}$ |
| Cf | S | FSH | Fracture specific heat capacity | $800 \frac{J}{kg \times K} - 1000 \frac{J}{kg \times K}$ |
| Ti | T | TI | Initial temperature | 150 °C – 200 °C |
| α | U | BC | Biot coefficient | 0.5 – 0.7 |
| β | V | TE | Thermal expansion coefficient | $10^{-6} \frac{1}{K} - 10^{-5} \frac{1}{K}$ |
| Tj | W | InjT | Injection temperature | 40 °C – 70 °C |

To understand the thermo-hydro-mechanical processes during a geothermal energy extraction operation, dynamics of thermophysical properties of $CO_2$ and geomechanical alteration needs attention. To understand the thermophysical properties during the operation period, we have plotted temperature, specific heat capacity of $CO_2$ at constant pressure and the dynamic viscosity of $CO_2$ in Figure 4 (a), (b) and (c) respectively. Since we assumed that the reservoir was fully saturated with hot $CO_2$, the temperature distribution after one year in Figure 4(a1) shows uniform cold front evolution around the injection well. However, temperature after 14 years in Figure 4(a2) indicates a directed flow towards the production well and it is well supported by the specific heat capacity distribution of $CO_2$ at constant pressure as shown in Figure 4(b1-b2). The production well temperature drops by 10 °C after 14 years whereas it takes 15.5 and 16.8 years to drop by 20 and 30 °C from the initial reservoir temperature, respectively. From the temperature distribution contour, it is clear that the presence of fracture connectivities allows the uniform evolution of colder front even after 14 years. Figure 4(c1) and 4(c2) shows viscosity of $CO_2$ at time one year and 14 years respectively. The viscosity decrease due to cooling of the reservoir, increase the flow rate and a rapid decrease in production well is observed. Approximately 40% of decrease in $CO_2$ viscosity reduction has negligible impact on directional flow of $CO_2$ along the fractures.

We have coupled hydraulic and thermal processes with geomechanical stresses in this work. We coupled thermoelasticity and poroelasticity to estimate the impact of cold $CO_2$ flow through the fractures and the rock matrix. Figure 5 demonstrates the poroelastic stress distribution in the reservoir. The horizontal and vertical components of poroelastic stresses are localized near the injection and production well. This localization is due to continuous injection of high density $CO_2$ at high pressure in high permeability fracture surrounded by low permeability rock matrix. The compressive stress near the production well is greater in the horizontal direction than vertical direction due to higher initial vertical stress compared to horizontal stress.

The thermoelastic stress has much wider distribution in the reservoir compared to poroelastic stress as shown in Figure 6. In this case, $CO_2$ is injected at 70 °C in a reservoir with initial temperature of 150 °C. The thermoelastic stress increases with decrease in specific heat capacity of fluid due to cold fluid spread in the reservoir. Poroelastic stress is highly localized in the vicinity of the injection-production wells whereas the thermal stress spreads along the cold fluid spread in the reservoir (see figure 6(a2 & b2)).

**3.2 Plackett- Burman methodology**

In this section, we have used design of experiments methodology to gather as much as information with least number of numerical simulations (Decaestecker et al. 2004, Mahmoodpour and Rostami 2017). Plackett-Burman methodology is one such technique that assists to identify the critical physical parameters with least simulations to run (Plackett and Burman 1946, Yannis 2001). Detailed information about methodology adopted in this work is discussed in our companion paper (Mahmoodpour et al. 2021). Using the Plackett-Burman method, we performed 48 numerical simulations in COMSOL for 22 input parameters (see Table 3) and identified relatively most important parameters among them. Table 4 list 48 numerical simulations with input values given in Table 3.

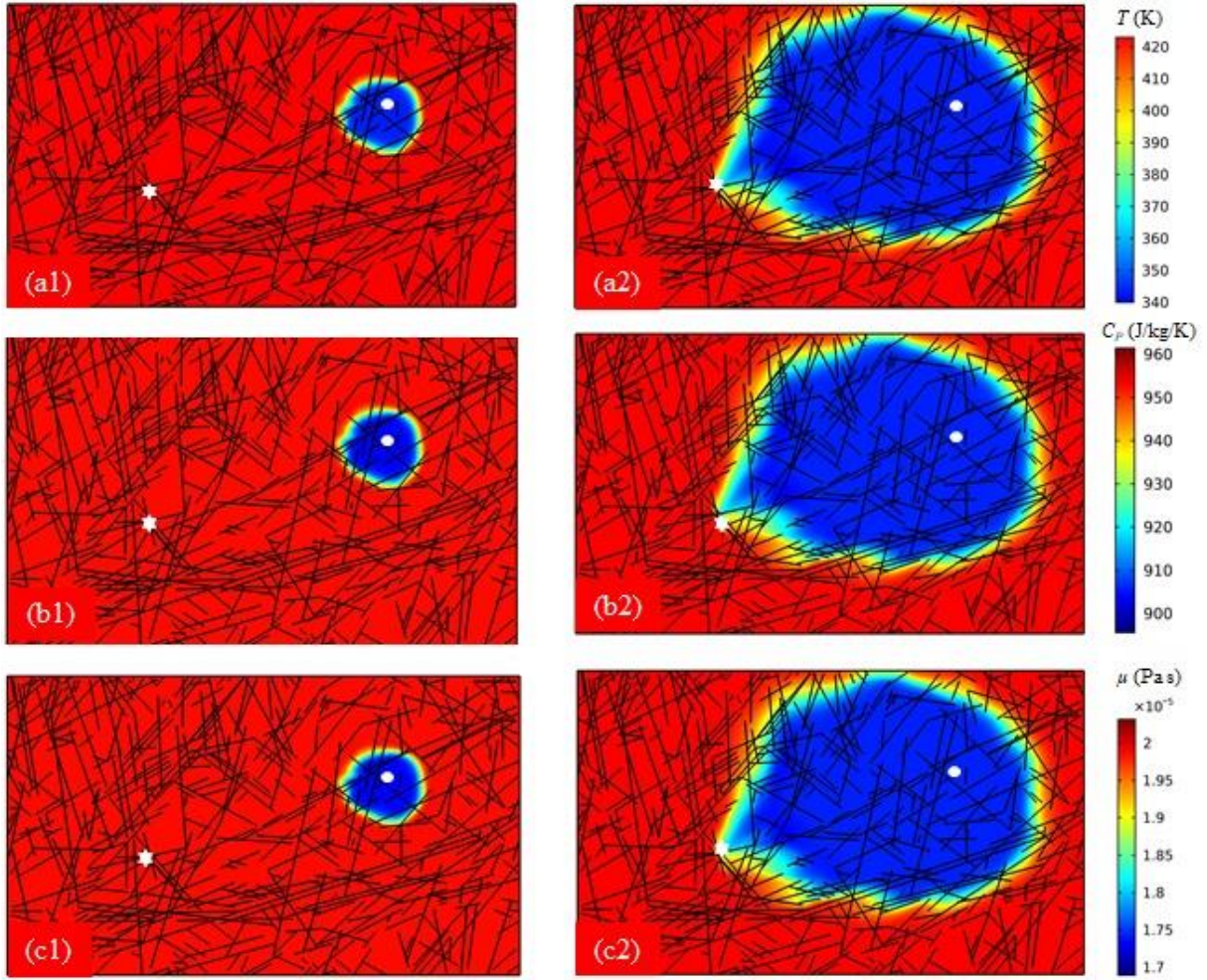

*Figure 4: (a) Fractured reservoir temperature, (b) specific heat capacity, and (c) dynamic fluid viscosity at (1) one and (2) 14 years. Here, the thermal breakthrough time at the production well is after 14 years. Here, white circle is injection well and white star is production well.*

Figures 7 and 8 demonstrate the results obtained from Pareto chart. The dash line with magnitude ±2.06 demarcates the important and unimportant parameters in both directions. The parameters in the positive direction indicates that they are directly influencing the final result whereas the negative direction shows that the final result is inversely proportional those parameters. We considered 5% level of uncertainty (α), indicating 95% confidence in our statistical estimation. We have shown the residual values between estimated and simulation results in Figures 9 & 10 for corresponding cases shown in Figures 7 & 8. From the residual plot we can observe that the residual follows a normal distribution (Figures 11 & 12).

### 3.3 Sensitivity study

In this section, we have presented the results obtained from the Plackett-Burman methodology sensitivity analysis. In the first stage, we measured the breakthrough time for temperature drop of 10 °C, 20 °C, and 30 °C at the production well for 48 cases. Similarly, we measured the thermal breakthrough time for 10% and 20% temperature drop at the production well. In the second stage, we estimated the mass flux and energy recovery at the production well for 10 °C, 20 °C, and 30 °C as well as 10% and 20% drop. Based on this data, we performed sensitivity analysis in the mass flux and energy recovery dependency on the input simulation parameters.

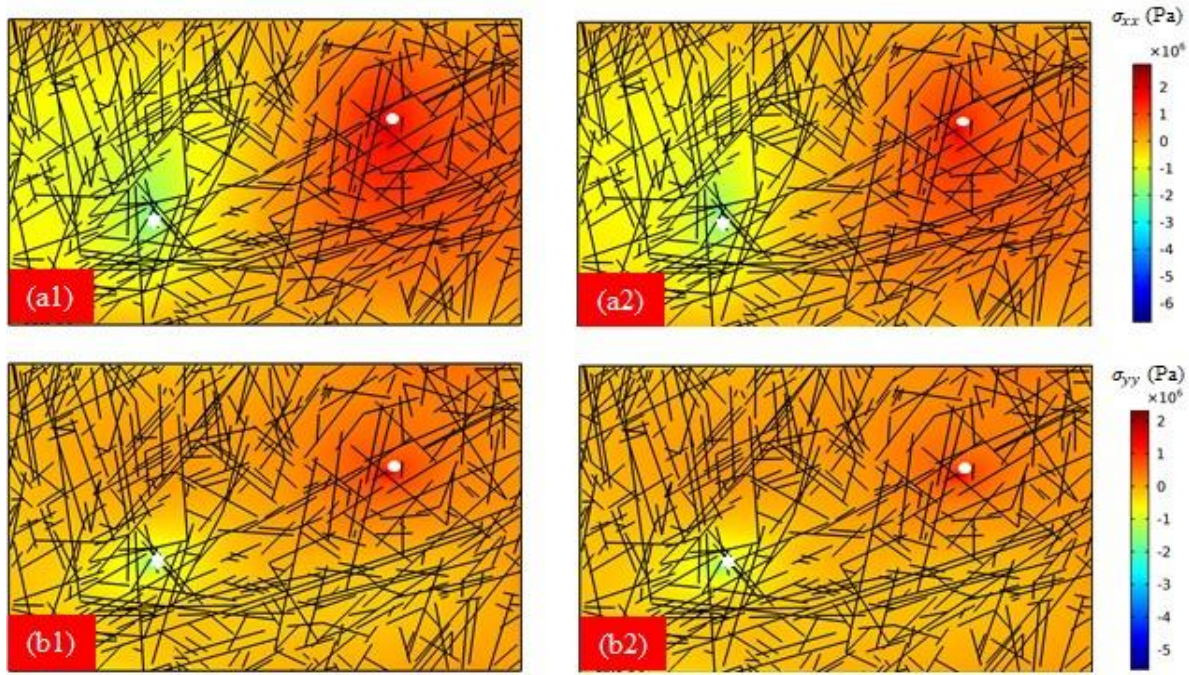

*Figure 5: Poroelastic stress distribution of (a) x-component and (b) y-component at time (1) one and (2) 14 years. Here, white circle is injection well and white star is production well.*

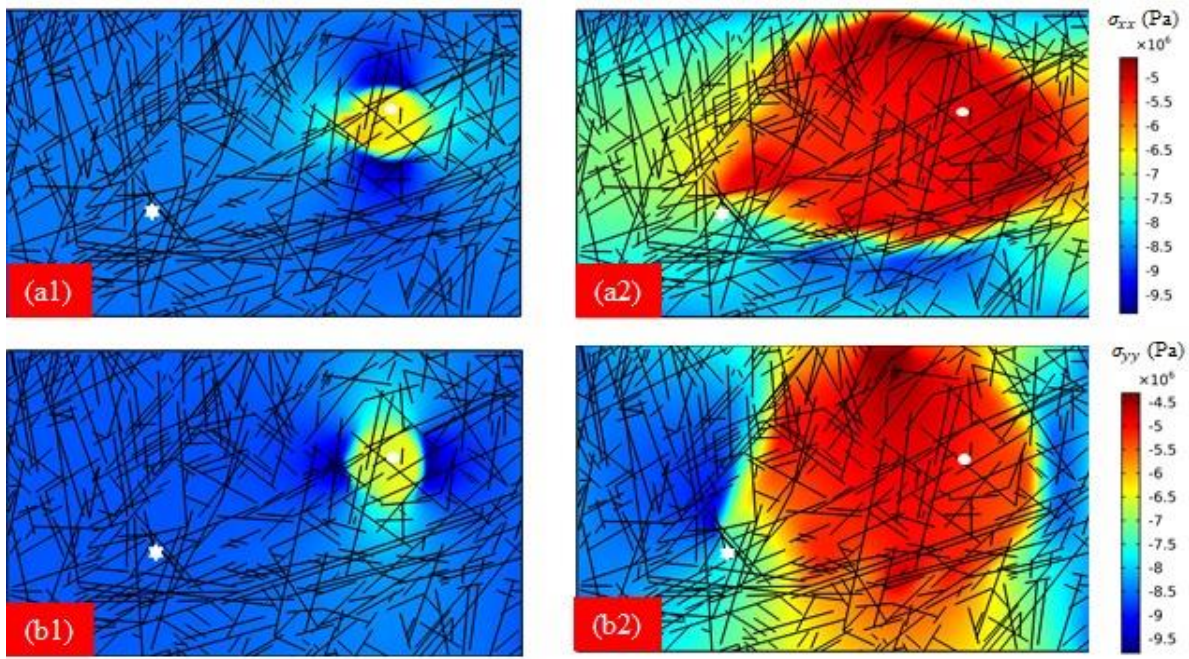

*Figure 6: Thermoelastic stress distribution of (a) x-component and (b) y-component at time (1) one and (2) 14 years. Here, white circle is injection well and white star is production well.*

*Table 4: List of 48 simulated cases (shown in rows) for 22 input parameters (shown by columns) used for performing the sensitivity study. This table was developed by using the Plackett-Burman method. The minimum and maximum values for parameters listed in Table 3. The minimal value is represented using -1 whereas maximum value is denoted by 1.*

| Case | E | ν | ρr | S1 | S2 | Pi | Pj | φr | kr | φf | f | Ap | σs | wr | λr | λf | Cr | Cf | Ti | α | β | Tj |
|---|---|---|---|---|---|---|---|---|---|---|---|---|---|---|---|---|---|---|---|---|---|---|
| 1 | 1 | -1 | -1 | -1 | 1 | 1 | -1 | 1 | 1 | -1 | -1 | 1 | -1 | -1 | 1 | 1 | 1 | -1 | 1 | -1 | 1 | -1 |
| 2 | 1 | -1 | -1 | -1 | -1 | 1 | -1 | -1 | -1 | -1 | 1 | 1 | -1 | 1 | -1 | 1 | -1 | -1 | -1 | 1 | 1 | -1 |
| 3 | -1 | -1 | 1 | -1 | -1 | 1 | 1 | 1 | -1 | 1 | -1 | 1 | -1 | -1 | 1 | 1 | 1 | 1 | -1 | 1 | 1 | 1 |
| 4 | -1 | -1 | -1 | -1 | 1 | -1 | -1 | -1 | -1 | 1 | 1 | -1 | 1 | -1 | 1 | -1 | -1 | -1 | 1 | 1 | -1 | 1 |
| 5 | -1 | -1 | 1 | 1 | 1 | -1 | 1 | -1 | 1 | -1 | -1 | 1 | 1 | 1 | 1 | -1 | 1 | 1 | 1 | 1 | 1 | -1 |
| 6 | 1 | 1 | -1 | -1 | 1 | -1 | -1 | 1 | 1 | 1 | -1 | 1 | -1 | 1 | -1 | -1 | 1 | 1 | 1 | 1 | -1 | 1 |
| 7 | -1 | -1 | -1 | -1 | 1 | 1 | -1 | 1 | -1 | 1 | -1 | 1 | -1 | -1 | 1 | -1 | 1 | 1 | -1 | -1 | 1 | -1 |
| 8 | 1 | 1 | 1 | 1 | -1 | 1 | 1 | 1 | 1 | 1 | -1 | -1 | -1 | -1 | 1 | -1 | -1 | -1 | -1 | 1 | 1 | -1 |
| 9 | -1 | 1 | -1 | -1 | 1 | 1 | 1 | -1 | 1 | -1 | 1 | -1 | -1 | 1 | 1 | 1 | 1 | -1 | 1 | 1 | 1 | 1 |
| 10 | -1 | 1 | 1 | -1 | 1 | -1 | 1 | -1 | -1 | -1 | 1 | 1 | -1 | 1 | 1 | -1 | -1 | 1 | -1 | -1 | 1 | 1 |
| 11 | -1 | 1 | 1 | 1 | -1 | 1 | -1 | 1 | -1 | -1 | 1 | 1 | 1 | 1 | -1 | 1 | 1 | 1 | 1 | 1 | -1 | -1 |
| 12 | 1 | 1 | -1 | -1 | -1 | -1 | 1 | -1 | -1 | -1 | -1 | 1 | 1 | -1 | 1 | -1 | 1 | -1 | -1 | -1 | 1 | 1 |
| 13 | 1 | 1 | -1 | 1 | -1 | 1 | -1 | -1 | 1 | 1 | 1 | 1 | -1 | 1 | 1 | 1 | 1 | 1 | -1 | -1 | -1 | -1 |
| 14 | 1 | 1 | 1 | 1 | 1 | -1 | -1 | -1 | -1 | 1 | -1 | -1 | -1 | 1 | 1 | -1 | 1 | -1 | 1 | 1 | -1 | -1 |
| 15 | 1 | -1 | -1 | 1 | 1 | 1 | 1 | -1 | 1 | 1 | 1 | 1 | 1 | -1 | -1 | -1 | -1 | 1 | -1 | -1 | -1 | -1 |
| 16 | -1 | 1 | -1 | 1 | -1 | -1 | -1 | 1 | 1 | 1 | 1 | -1 | -1 | 1 | -1 | -1 | 1 | 1 | 1 | -1 | 1 | 1 |
| 17 | 1 | 1 | -1 | 1 | 1 | -1 | -1 | 1 | -1 | -1 | 1 | 1 | 1 | -1 | 1 | -1 | 1 | -1 | -1 | 1 | 1 | 1 |
| 18 | 1 | -1 | 1 | -1 | 1 | -1 | -1 | -1 | 1 | 1 | -1 | 1 | -1 | -1 | 1 | -1 | -1 | 1 | 1 | 1 | 1 | -1 |
| 19 | -1 | -1 | 1 | 1 | -1 | 1 | -1 | 1 | -1 | -1 | -1 | 1 | 1 | -1 | 1 | 1 | -1 | -1 | 1 | -1 | -1 | 1 |
| 20 | -1 | 1 | 1 | -1 | -1 | 1 | -1 | -1 | 1 | 1 | 1 | -1 | 1 | -1 | 1 | -1 | -1 | 1 | 1 | 1 | 1 | -1 |
| 21 | 1 | -1 | 1 | -1 | -1 | -1 | 1 | 1 | -1 | 1 | 1 | -1 | -1 | 1 | -1 | -1 | 1 | 1 | 1 | -1 | 1 | -1 |
| 22 | 1 | -1 | -1 | 1 | 1 | 1 | -1 | 1 | -1 | 1 | -1 | -1 | 1 | 1 | 1 | 1 | -1 | 1 | 1 | 1 | 1 | 1 |
| 23 | -1 | 1 | 1 | 1 | 1 | -1 | 1 | 1 | 1 | 1 | 1 | -1 | -1 | -1 | -1 | 1 | -1 | -1 | -1 | -1 | 1 | 1 |
| 24 | 1 | 1 | 1 | -1 | 1 | -1 | 1 | -1 | -1 | 1 | 1 | 1 | 1 | -1 | 1 | 1 | 1 | 1 | 1 | -1 | -1 | -1 |
| 25 | 1 | 1 | 1 | -1 | -1 | -1 | -1 | 1 | -1 | -1 | -1 | -1 | 1 | 1 | -1 | 1 | -1 | 1 | -1 | -1 | -1 | 1 |
| 26 | 1 | -1 | 1 | 1 | 1 | 1 | 1 | -1 | -1 | -1 | -1 | 1 | -1 | -1 | 1 | -1 | 1 | 1 | -1 | 1 | -1 | 1 |
| 27 | 1 | -1 | 1 | 1 | -1 | -1 | 1 | -1 | -1 | 1 | 1 | -1 | 1 | -1 | 1 | -1 | -1 | 1 | 1 | 1 | 1 | 1 |
| 28 | -1 | -1 | 1 | 1 | 1 | 1 | -1 | 1 | 1 | 1 | 1 | 1 | -1 | -1 | -1 | -1 | 1 | -1 | -1 | -1 | -1 | 1 |
| 29 | -1 | -1 | -1 | -1 | -1 | -1 | -1 | -1 | -1 | -1 | -1 | -1 | -1 | -1 | -1 | -1 | -1 | -1 | -1 | -1 | -1 | -1 |
| 30 | 1 | 1 | -1 | 1 | 1 | 1 | 1 | 1 | -1 | -1 | -1 | -1 | 1 | -1 | -1 | -1 | 1 | 1 | -1 | 1 | -1 | -1 |
| 31 | -1 | -1 | -1 | 1 | 1 | -1 | 1 | -1 | 1 | -1 | -1 | -1 | 1 | -1 | 1 | 1 | -1 | -1 | 1 | -1 | -1 | -1 |
| 32 | -1 | -1 | 1 | 1 | -1 | 1 | 1 | -1 | 1 | 1 | -1 | -1 | 1 | 1 | -1 | 1 | -1 | 1 | -1 | -1 | -1 | 1 |
| 33 | -1 | -1 | -1 | 1 | -1 | -1 | -1 | -1 | 1 | 1 | 1 | -1 | 1 | -1 | -1 | -1 | -1 | 1 | 1 | -1 | 1 | 1 |
| 34 | -1 | 1 | 1 | 1 | 1 | 1 | -1 | -1 | -1 | 1 | -1 | -1 | -1 | -1 | 1 | 1 | -1 | 1 | -1 | 1 | -1 | -1 |
| 35 | 1 | 1 | 1 | 1 | -1 | -1 | -1 | -1 | 1 | -1 | -1 | -1 | -1 | 1 | 1 | -1 | 1 | -1 | 1 | -1 | -1 | -1 |
| 36 | 1 | -1 | 1 | -1 | 1 | -1 | -1 | 1 | 1 | 1 | 1 | -1 | 1 | 1 | 1 | 1 | -1 | -1 | -1 | -1 | -1 | 1 |
| 37 | 1 | -1 | 1 | -1 | -1 | 1 | 1 | 1 | 1 | -1 | 1 | 1 | 1 | 1 | -1 | -1 | -1 | -1 | 1 | -1 | 1 | -1 |
| 38 | -1 | -1 | -1 | 1 | 1 | -1 | 1 | 1 | -1 | -1 | 1 | -1 | 1 | 1 | 1 | -1 | 1 | -1 | 1 | -1 | -1 | -1 |
| 39 | -1 | 1 | 1 | -1 | 1 | 1 | -1 | 1 | -1 | -1 | 1 | 1 | 1 | -1 | 1 | -1 | 1 | -1 | -1 | 1 | 1 | 1 |
| 40 | -1 | -1 | 1 | -1 | -1 | -1 | 1 | 1 | 1 | 1 | -1 | 1 | -1 | -1 | -1 | 1 | 1 | -1 | 1 | 1 | 1 | -1 |
| 41 | -1 | 1 | -1 | 1 | -1 | -1 | 1 | 1 | 1 | 1 | -1 | 1 | 1 | 1 | 1 | -1 | -1 | -1 | -1 | 1 | -1 | -1 |
| 42 | 1 | 1 | 1 | -1 | 1 | 1 | 1 | 1 | 1 | -1 | -1 | -1 | -1 | 1 | -1 | -1 | -1 | -1 | 1 | 1 | -1 | 1 |
| 43 | -1 | 1 | -1 | -1 | -1 | -1 | 1 | 1 | -1 | 1 | -1 | 1 | -1 | -1 | -1 | 1 | 1 | -1 | 1 | 1 | -1 | -1 |

| 44 | 1 | -1 | -1 | -1 | -1 | 1 | 1 | -1 | 1 | -1 | 1 | -1 | -1 | -1 | 1 | 1 | -1 | 1 | 1 | -1 | -1 | 1 |
| 45 | -1 | 1 | -1 | -1 | -1 | 1 | 1 | -1 | 1 | 1 | -1 | -1 | 1 | -1 | -1 | 1 | 1 | 1 | -1 | 1 | -1 | 1 |
| 46 | 1 | 1 | -1 | 1 | -1 | 1 | -1 | -1 | -1 | 1 | 1 | -1 | 1 | 1 | -1 | -1 | 1 | -1 | -1 | 1 | 1 | 1 |
| 47 | -1 | 1 | -1 | -1 | 1 | 1 | 1 | 1 | -1 | 1 | 1 | 1 | 1 | 1 | -1 | -1 | -1 | -1 | 1 | -1 | -1 | -1 |
| 48 | 1 | -1 | -1 | 1 | -1 | -1 | 1 | 1 | 1 | -1 | 1 | -1 | 1 | -1 | -1 | 1 | 1 | 1 | 1 | -1 | 1 | 1 |

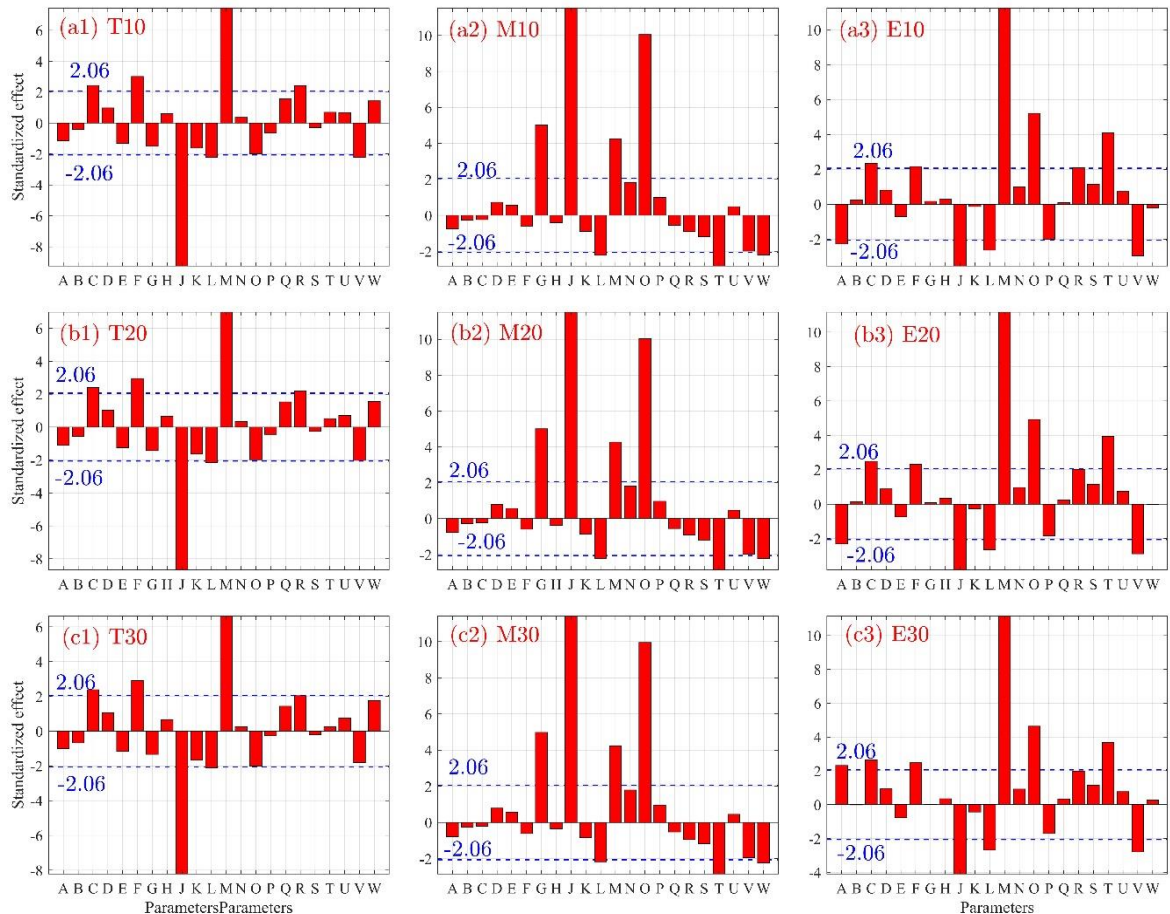

*Figure 7: Pareto charts of 22 parameters to investigate the comparative impact on THM model for the production well breakthrough temperature. The list of 22 parameters is given in table 5 with their corresponding symbols. The left column represent the temperature breakthrough sensitivity for a (a1) 10 ℃, (b1) 20 ℃, and (c1) 30 ℃ temperature drop, the middle column shows the sensitivity in the mass flow rate at the production well with a (a2) 10 ℃, (b2) 20 ℃, and (c2) 30 ℃ drop in the well temperature, and the right column indicates the energy sensitivity for corresponding the temperature drop by (a3) 10 ℃, (b3) 20 ℃, and (c3) 30 ℃. Here, T10 means the temperature sensitivity study for a 10 ℃ temperature drop. Similarly, M10 and E10 indicates a sensitivity for a temperature drop at the production well for 10 ℃, respectively.*

Results obtained from the Plackett-Burman methodology is shown using the Pareto chart in Figure 7. The sensitivity analysis for temperature breakthrough, mass flux and total energy extraction are performed using 48 numerical simulations and shown in Figure 7(a1, b1 & c1), 7(a2, b2 & c2) and 7(a3, b3 & c3), respectively. In the first step, breakthrough time for a temperature drop of 10 ℃, 20 ℃ and 30 ℃ at the production well is estimated. Based on this thermal breakthrough time, sensitivity in mass flux and total energy recovery are performed. Figure 9 and Figure 11 shows the residual and histogram plots corresponding to the cases shown in Figure 7. From this sensitivity analysis, we find that the key parameter governing the thermal breakthrough time during the THM process for energy extraction using $CO_2$ as heat transmitting fluid are matrix permeability and fracture aperture (see Figure 7(a1, b1 and c1)). For the given range of matrix permeability, thermal breakthrough time is small for high permeability whereas thermal breakthrough time is higher for higher fracture aperture. Higher matrix permeability causes faster fluid flow in the matrix and therefore, thermal depletion rate reduces. On the other hand, larger fracture aperture increases the fluid velocity in the fracture and therefore, heat exchange time between fluid and rock matrix becomes smaller leading to multiple short-circuiting in along the fractures. Hence, higher fracture aperture results in faster thermal breakthrough time. The matrix permeability and fracture aperture are equally important in deciding the thermal breakthrough time as shown in Figure 7(a1, b1 & c1). In comparison to thermal breakthrough time, a large set of parameters critically controls the mass flux sensitivity (see Figure 7(a2, b2 and c2)). The important among these parameters are matrix permeability and wellbore radius and they are important by a factor of approximately two with respect to injection pressure and fracture aperture. It is worth to note that initial reservoir temperature also plays an important role in deciding the mass flux at the production well. However, initial reservoir temperature has almost five times less impact on mass flux when compared to matrix permeability. Further, sensitivity analysis results for total enery recovery shows a trend similar to thermal breakthrough time (see Figure 7(a3, b3 and c3)). However, energy recovery indicates approximately four times greater dependency on fracture aperture than matrix permeability as heat mining is greatly controlled by flow occurring through the fractures. Since mass flux at the production well is directly linked with total energy extraction, increase in wellbore radius and initial reservoir temperature enhances the total energy recovery. The absolute magnitude of residuals between the simulated and regression results for all 48 cases corresponding to Figure 7 follows a normal distribution (See Figure 11).

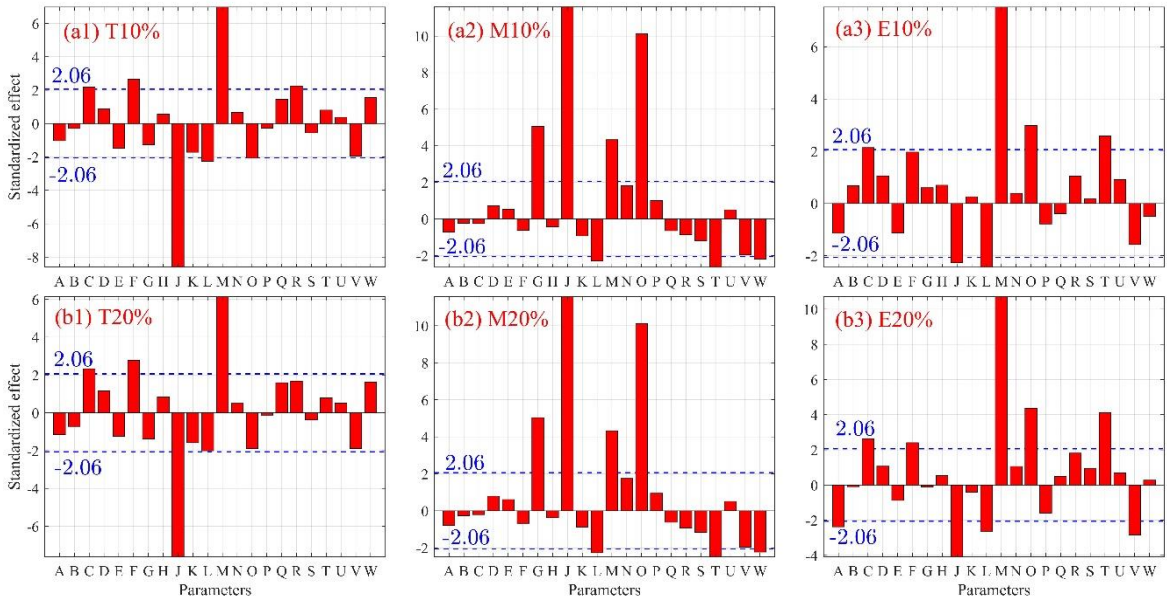

*Figure 8: Pareto charts of 22 parameters to investigate the comparative impact on THM model for the production well breakthrough temperature. The list of 22 parameters is given in table 5. The left column represents the temperature breakthrough sensitivity for a (a1) 10% and (b1) 20% temperature drop, the middle column shows the sensitivity in the mass flow rate at the production well for a (a2) 10% and (b2) 20% drop in the well temperature, and the right column indicates the energy sensitivity for the corresponding temperature drop by (a3) 10% and (b3) 20%.*

In the next step, sensitivity analysis for a 10% and 20% drop in temperature at the production well are shown in Figure 8(a1, a2 & a3) and 8(b1, b2 & b3) respectively. Correspondingly, residuals between simulation and regression values and their distribution are shown in Figure 10 and 12 respectively. Except for the total energy recovery, thermal breakthrough time and mass flux sensitivity results shows similar parameter dependency as shown in Figure 7. This is due to amplified dependency of energy recovery on fracture aperture.

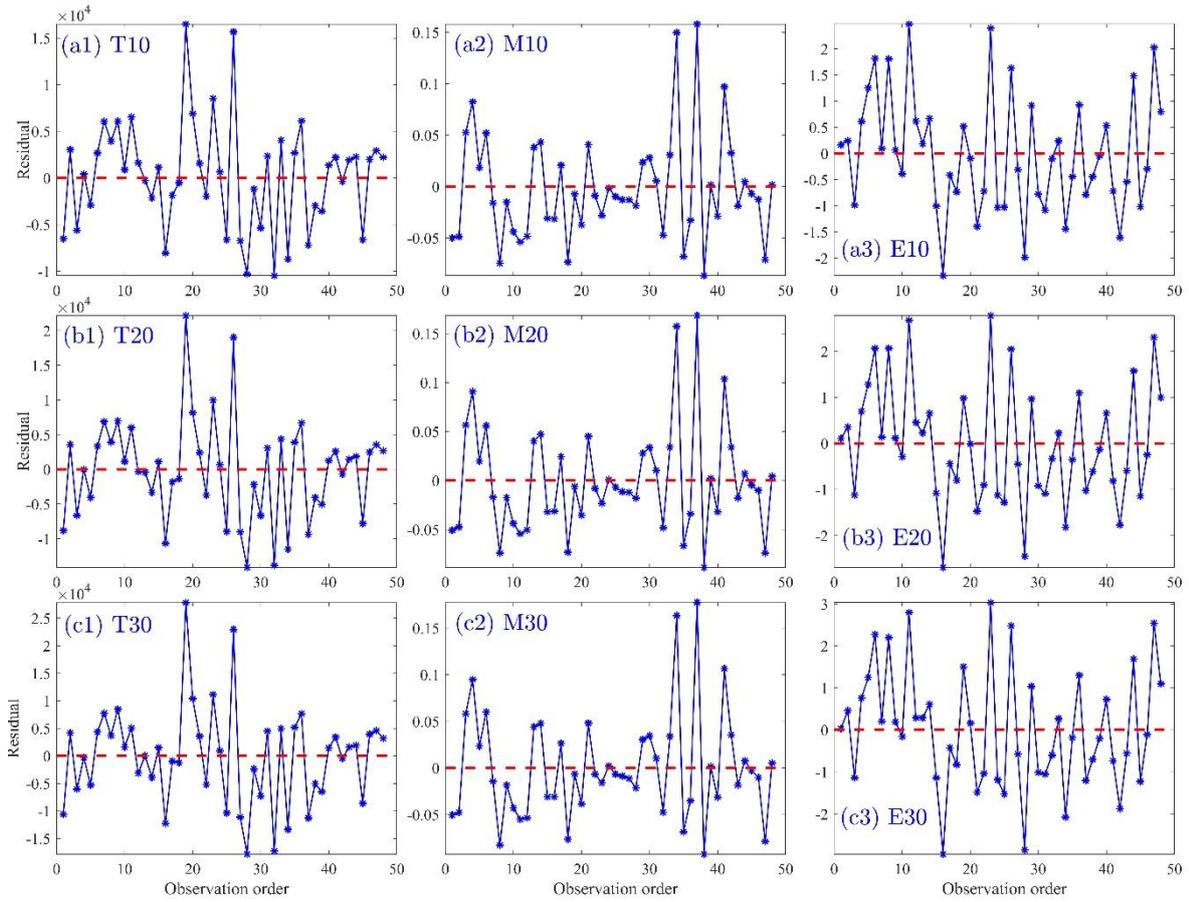

*Figure 9: Residual between simulated and regression results for 48 cases. The left column represents the temperature breakthrough sensitivity for a (a1) 10 ℃, (b1) 20 ℃, and (c1) 30 ℃ temperature drop, the middle column shows the sensitivity in the mass flow rate at the production well for a (a2) 10 ℃, (b2) 20 ℃, and (c2) 30 ℃ drop in the well temperature, and the right column indicates the energy sensitivity for the corresponding temperature drop by (a3) 10 ℃, (b3) 20 ℃, and (c3) 30 ℃.*

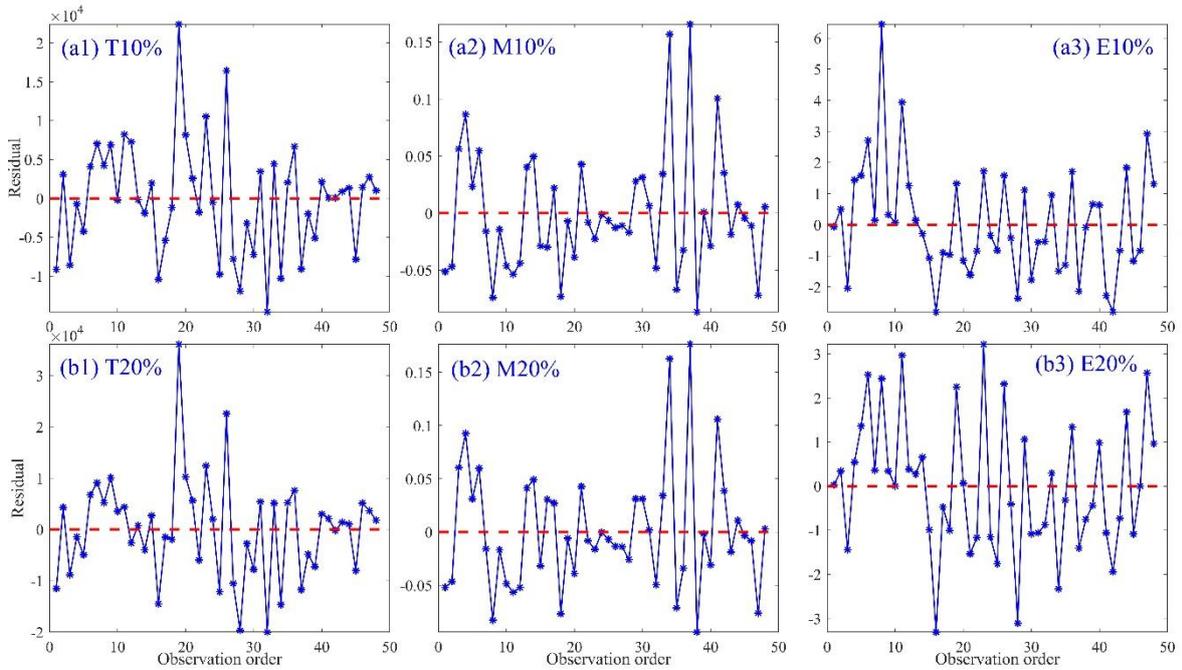

*Figure 10: Residual between simulated and regression results for 48 cases. The left column represents the temperature breakthrough sensitivity for a (a1) 10% and (b1) 20% temperature drop, the middle column shows the sensitivity in the mass flow rate at the production well for a (a2) 10% and (b2) 20% drop in the well temperature, and the right column indicates the energy sensitivity for the corresponding temperature drop by (a3) 10% and (b3) 20%.*

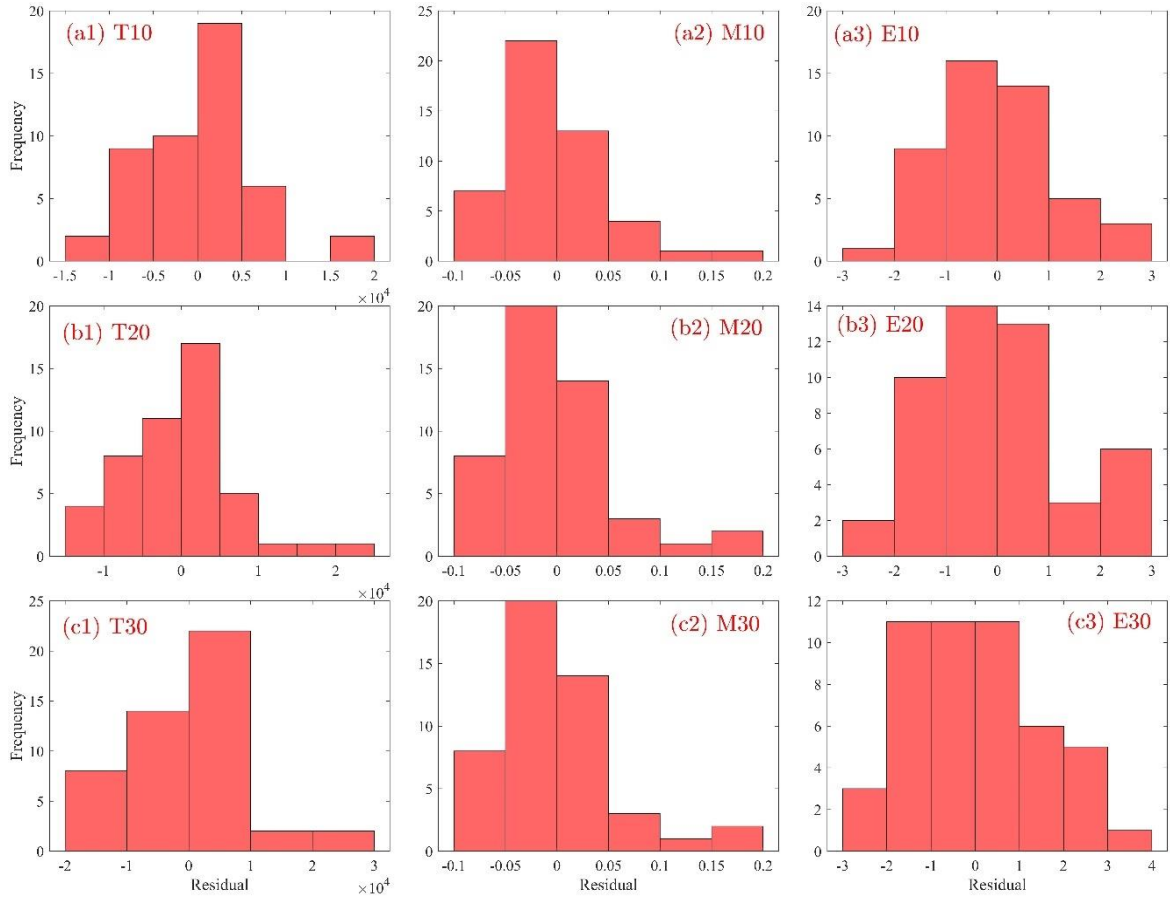

*Figure 11: Histogram of residuals between simulated and regression results. The left column represents the temperature breakthrough sensitivity for a (a1) 10 ℃, (b1) 20 ℃, and (c1) 30 ℃ temperature drop, the middle column shows the sensitivity in the mass flow rate at the production well for a (a2) 10 ℃, (b2) 20 ℃, and (c2) 30 ℃ drop in the well temperature, and the right column indicates the energy sensitivity for the corresponding temperature drop by (a3) 10 ℃, (b3) 20 ℃, and (c3) 30 ℃.*

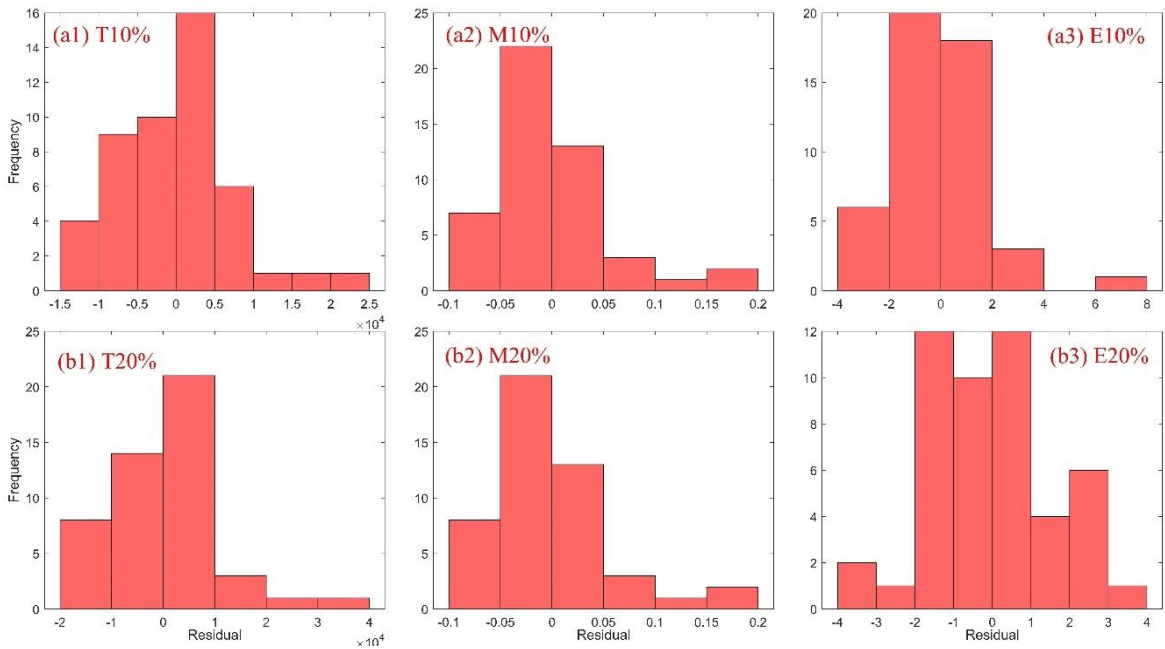

*Figure 12: Histogram of residuals between simulated and regression results for 48 cases. The left column represents the temperature breakthrough sensitivity for a (a1) 10% and (b1) 20% temperature drop, the middle column shows the sensitivity in the mass flow rate at the production well for a (a2) 10% and (b2) 20% drop in the well temperature, and right column indicates the energy sensitivity for the corresponding temperature drop by (a3) 10% and (b3) 20%.*

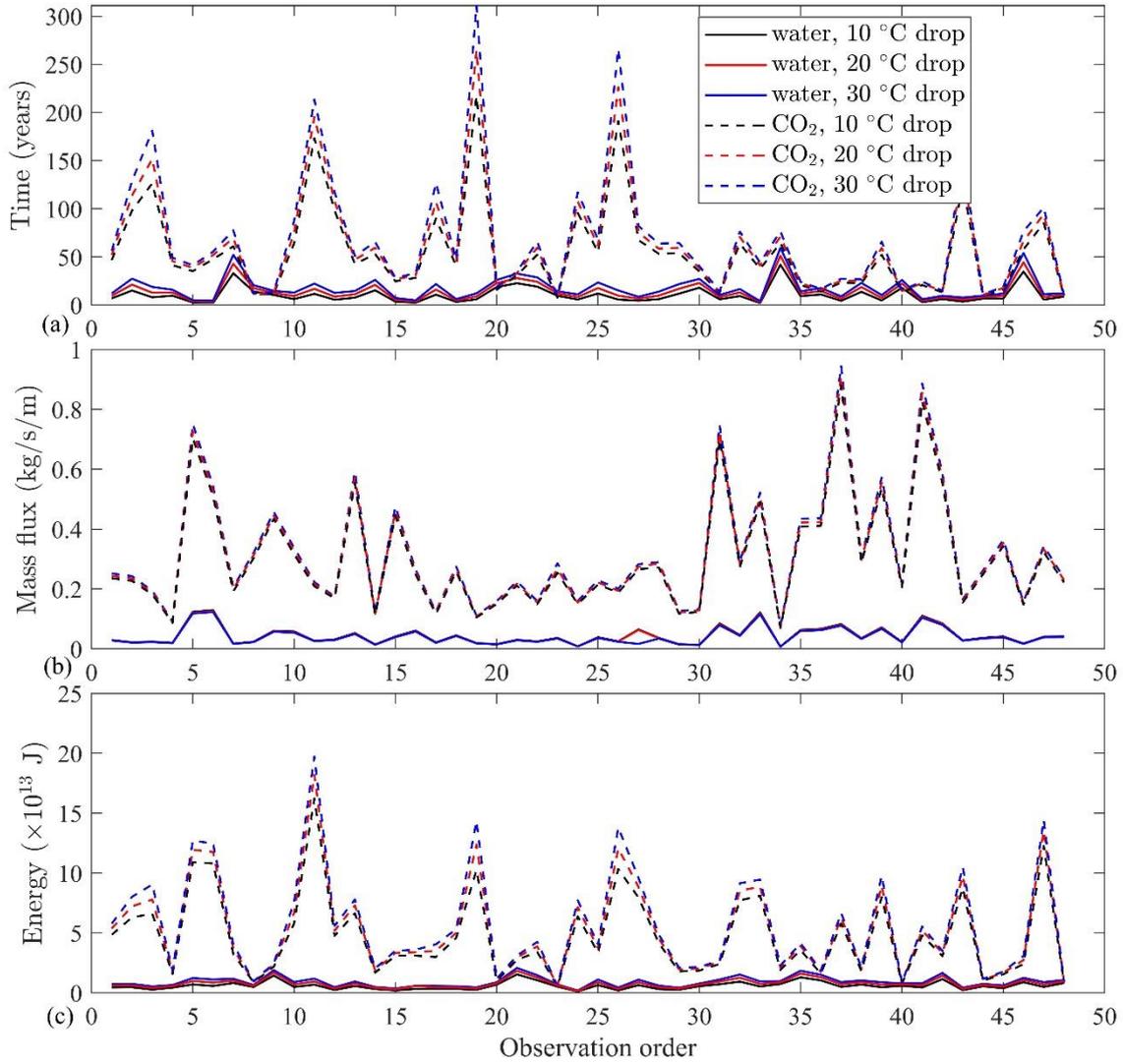

*Figure 13: (a) Comparison of breakthrough time between water and CO₂ geothermal energy extraction system for a temperature drop of 10 ℃, 20 ℃ and 30 ℃ at the production well, (b) comparison of mass flux at the production well between water and CO₂ geothermal energy extraction system for a temperature drop of 10 ℃, 20 ℃ and 30 ℃ at the production well, and (c) comparison of total energy recovered at the production well between water and CO₂ geothermal energy extraction system for a temperature drop of 10 ℃, 20 ℃ and 30 ℃ at the production well.*

Now, we will compare the results obtained from sensitivity analysis for geothermal energy extraction operation for two different fluids: water (Mahmoodpour et al. 2021) and $CO_2$. Even though, our model input parameters and boundary conditions are approximately same with that of Mahmoodpour et al. (2021), significantly different results are noticeable in this study. For water-EGS cases, fracture aperture is the most important parameter whereas for $CO_2$-EGS, matrix permeability and fracture aperture are more influencial. Figure 13 shows the comparison between water-EGS and $CO_2$-EGS for thermal breakthrough time, mass flux and energy recovery all three at the production well for three different temperature drops and 48 numerical simulations. The key difference between water and $CO_2$ result is the thermal breakthrough time. Approximately one order faster thermal breakthrough happens for water-EGS due to higher heat capacity as shown in Figure 13(a). Due to this higher breakthrough time, greater amount of hot fluid production is achievable at the production well and concomitantly, higher total energy recovery is possible as shown in Figure 13(b) and 13(c) respectively.

## 4. Conclusions

Geothermal energy extraction using $CO_2$ as heat carrying medium can serve two purposes: heat mining from fractured reservoir and $CO_2$ geosequestration. This study focuses on heat extraction from a discretely fractured reservoir and demonstrates the invloved thermo-hydro-mechanical processes. There are no existing sensitivity analysis for EGS operation when the heat transfer fluid is $CO_2$. This study fills that gap and verified key parameters governing the THM process. As part of the MEET project, the objective of this study is to present a detailed and comprehensive sensitivity analysis for future prospective usage of $CO_2$ for heat extraction from the fractured reservoirs. A fully coupled poroelastic and thermoelastic geomechanical stress conditions were implemented for a two-dimensional fractured reservoir. The THM numerical simulations were further used to perform sensitivity analysis using Plackett-Burman methodology and identify critical parameters among 22 input parameters. We simulated 48 numerical simulations to feed input for the sensitivity analysis on thermal breakthrough time, mass flux and the total energy recovery from the reservoir. We have found that matrix permeability and fracture aperture are the most important parameter governing the thermal breakthrough time at the production well. Mass flux is primarily governed by matrix permeability, wellbore radius, injection pressure and the fracture aperture whereas total energy recovery is controlled by fracture aperture, wellbore radius, initial matrix temperature and matrix permeability. Further, we have demonstrated that residuals of the sensitivity analysis follows normal distribution and therefore, they estimate the relative significance of the involved parameter exhaustively.

**Acknowledgement**

This work is funded by European Union's Horizon 2020 research and innovation programme under grant agreement No 792037 for the MEET project. Group of Geothermal Science and Technology, Institute of Applied Geosciences, Technische Universität Darmstadt has provided institutional support to authors.

| Symbol | Parameter |
|---|---|
| $p$ | Fluid pressure |
| $T$ | Fluid Temperature |
| $\varepsilon_V$ | Pore volumetric strain |
| $\alpha_m$ | Biot's coefficient of porous media |
| $\alpha_f$ | Biot's coefficient of the fracture |
| $\phi_m$ | Reservoir porosity |
| $\phi_f$ | Fracture zone porosity |
| $S_m$ | Storage coefficients of fluid |
| $S_1$ | Storage coefficients of rock matrix |
| $S_f$ | Storage coefficients of fracture |
| $\beta_1$ | Thermal expansion coefficients of fluid |
| $\beta_m$ | Thermal expansion coefficients of rock matrix |
| $\beta_f$ | Thermal expansion coefficient of fracture |
| $\rho \, \& \, \rho_1$ | Fluid density |
| $k_m$ | pressure-dependent rock matrix permeability |
| $k_f$ | stress-dependent fracture permeability |
| $e_h$ | hydraulic aperture between two fracture surfaces |
| $nQ_m$ | $n.(-\frac{\rho k_m}{\mu \nabla p})$, mass flux exchange between porous media and the fracture |

| $\nabla_T$ | Gradient operator restricted to the fracture's tangential plane |
|---|---|
| $T_m$ | Rock matrix temperature |
| $T_l$ | Fluid temperature |
| $\rho_m$ | Rock density |
| $C_{p,m}$ | Specific heat capacity of the rock matrix |
| $\lambda_m$ | Heat conductivity of the rock matrix |
| $q_{ml}$ | Rock matrix-pore fluid interface heat transfer coefficient |
| $\rho_f$ | density of the fracture zone |
| $C_{p,f}$ | Specific heat capacity of the fracture |
| $\lambda_f$ | Heat conductivity of the fracture |
| $q_{fl}$ | Rock fracture-fluid interface heat transfer coefficient |
| $C_p$ & $C_{p,l}$ | Heat capacity of the fluid at a constant pressure |
| $\lambda_l$ | Heat conductivity of the fluid |
| $\sigma_{ij}$ | Total stress |
| G & $\lambda$ | Lame's constants |
| tr | Trace operator |
| $K'$ | $\frac{2G(1+\nu)}{3(1-2\nu)}$, bulk modulus of the drained porous media |
| $\beta_T$ | Volumetric thermal expansion coefficient of porous media |
| $\delta_{ij}$ | Dirac dealt function |
| $\alpha_p$ | Biot's coefficient |
| $\sigma_{eff}^{ij}$ | Effective stress |
| $f_i$ | External body force |
| $\Delta e_n$ | Change in the initial aperture of the fracture under in-situ stresses |
| $e_0$ | Initial aperture of the fracture |
| $\sigma_{eff}^n$ | Effective normal stress acting on the fracture surface |
| $\sigma_{nref}$ | Effective normal stress required to cause 90% reduction in fracture aperture |
| $\mu$ | $CO_2$ dynamic viscosity |
| $\kappa$ | $CO_2$ thermal conductivity |

**Reference**

1. Bai, B., "One-dimensional thermal consolidation characteristics of geotechnical media under non-isothermal condition", Eng. Mech., 22, p: 186-191, 2005.
2. Bai B., He Y., Li X., "Numerical study on the heat transfer characteristics between supercritical carbon dioxide and granite fracture wall", Geothermics, 2018, 75:40–47.
3. Bandis, S, C., Lumsden, A, C., Barton, N, R., "Fundamentals of rock joint deformation", International Journal of Rock Mechanics and Mining Sciences & Geomechanics Abstracts, 20, 6, p:249-268, 1983.
4. Barton, N., Bandis, S., Bakhtar, K., "Strength, deformation and conductivity coupling of rock joints", International Journal of Rock Mechanics and Mining Sciences & Geomechanics Abstracts, 22, 3, p: 121-140, 1985.
5. Brown D., "A hot dry rock geothermal energy concept utilizing supercritical $CO_2$ instead of water", In 25th workshop on geothermal reservoir engineering, 2000, Stanford University, Stanford. SGP-TR-165.
6. Chen Y., Ma G., Wang H., Li T., Wang Y., "Application of carbon dioxide as working fluid in geothermal development considering a complex fractured system", Energ Convers Manag, 2019, 180:1055–1067.
7. COMSOL Multiphysics® v. 5.5. www.comsol.com. COMSOL AB, Stockholm, Sweden.

8. Decaestecker TN, Lambert WE, Peteghem Carlos HV, Deforce D, Van Bocxlaer JF. Optimization of solid-phase extraction for a liquid chromatographic-tandem mass spectrometric general unknown screening procedure by means of computational techniques. J Chrom A 2004; 1056: 57–65.
9. Guo T., Gong F., Wang X., Lin Q., Qu Z., Zhang W., "Performance of enhanced geothermal system (EGS) in fractured geothermal reservoirs with $CO_2$ as working fluid", Appl Therm Eng, 2019, 152:215–230.
10. IPCC, "IPCC special report on carbon dioxide capture and storage", 2005, Cambridge University Press, New York.
11. Liao, J., Cao, C., Hou, Z., Mehmood, F., Feng, W., Yue, Y., Liu, H., "Field scale numerical modeling of heat extraction in geothermal reservoir based on fracture network creation with supercritical $CO_2$ as working fluid", Environ. Earth Sci. 2020, 79, 291.
12. Liu L., Suto Y., Bignall G., Yamasaki N., Hashida T., "$CO_2$ injection to granite and sandstone in experimental rock/hot water systems", Energy Convers Manag, 2003, 44(9):1399–1410.
13. Liu Y., Wang G., Yue G., Lu C., Zhu X., "Impact of $CO_2$ injection rate on heat extraction at the HDR geothermal field of Zhacanggou", China. Environ Earth Sci, 2017, 76(6):1–11.
14. Luo F., Xu R.N., Jiang P.X.,"Numerical investigation of fluid flow and heat transfer in a doublet enhanced geothermal system with $CO_2$ as the working fluid ($CO_2$-EGS)", Energy, 2014, 64:307–322.
15. Mahmoodpour, S., Masihi, M., "An improved simulated annealing algorithm in fracture network modeling", Journal of Natural Gas Science and Engineering, 33, p: 538-550, 2016.
16. Mahmoodpour, S., Rostami, B., "Design-of-experiment-based proxy models for the estimation of the amount of dissolved $CO_2$ in brine: A tool for screening of candidate aquifers in geo-sequestration", Int. J. Green. Gas Cont. 2017, 56, 261-277.
17. Mahmoodpour, S., Rostami, B., Soltanian, M.R., Amooie, A., "Convective dissolution of carbon dioxide in deep saline aquifers: Insights from engineering a high-pressure porous cell", Phys. Rev. Appl. 2019, 12, 034016.
18. Mahmoodpour, S., Singh, M., Turan, A., Bär, K., Sass, I., "Key parameters affecting the performance of fractured geothermal reservoirs: a sensitivity analysis by thermo-hydraulic-mechanical simulation", http://arxiv.org/abs/2107.02277.
19. Plackett, R, L., Burman, J, P., "The design of optimum multifactorial experiments", Biometrika, 33 (4), p: 305-325, 1946.
20. Pruess K., "Enhanced geothermal systems ( EGS ) using $CO_2$ as working fluid - a novel approach for generating renewable energy with simultaneous sequestration of carbon", Geothermics, 2006, 35:351–367.
21. Pruess K., "On production behavior of enhanced geothermal systems with $CO_2$ as working fluid", Energy Convers Manag, 2008, 49(6):1446–1454.
22. Shao S., Wasantha P., Ranjith P., Chen B., "Effect of cooling rate on the mechanical behavior of heated strathbogie granite with different grain sizes", Int J Rock Mech Min Sci, 2014, 70:381–387.
23. Singh, M., Chaudhuri, A., Stauffer, P.H., Pawar, R.J., "Simulation of gravitational instability and thermo-solutal convection during the dissolution of $CO_2$ in deep storage reservoirs", Wat. Resour. Resear. 2020a, 56, e2019WR026126.
24. Singh, M., Tangirala, S.K., Chaudhuri, A., "Potential of $CO_2$ based geothermal energy extraction from hot sedimentary and dry rock reservoirs, and enabling carbon geo-sequestration", Geom. Geop. Geo-Energy Geo-resour., 2020b, 6, 16.
25. Singh, M., Chaudhuri, A., Soltanian, M.R., Stauffer, P.H., "Coupled multiphase flow and transport simulation to model $CO_2$ dissolution and local capillary trapping in permeability and capillary heterogeneous reservoir", Int. J. Green. Gas. Cont. 2021, 108, 103329.
26. Sun, Z., Bongole, K., Yao, J., Wang, Y., Mboje, J., Seushi, K., Jiang, C., Wang, T., "Combination of double and single cyclic pressure alteration technique to increase $CO_2$ sequestration with heat mining in enhanced geothermal reservoirs by thermo-hydro-mechanical coupling methods", Int. J Energy Resear. 2019, 44, 3478-3496.
27. Tang Y., Ma T., Chen P., Ranjith P.G., "An analytical model for heat extraction through multi-link fractures of the enhanced geothermal system", Geomech Geophys Geo Energy Geo Res, 2019, 6(1):1.
28. Wang Y., Li T., Chen Y., Ma G., "Numerical analysis of heat mining and geological carbon sequestration in supercritical $CO_2$ circulating enhanced geothermal systems inlayed with complex discrete fracture networks", Energy, 2019, 173:92–108.
29. Yang, S.-Y., Yeh, H.-D., "Modeling heat extraction from hot dry rock in a multiwell system", Appl. Therm. Engg. 2009, 29 (8e9) 1676e1681.
30. Yannis L.L., "A Plackett-Burnam screening design directs the efficient formulation of multicomponent DRV liposomes", J Pharm Biomed Anal, 2001; 26: 255–263.
31. Zhang L., Luo F., Xu R., Jiang P., Liu H., "Heat transfer and fluid transport of supercritical $CO_2$ in enhanced geothermal system with local thermal non-equilibrium model", Energy Procedia, 2014, 63:7644–7650.
32. Zhang Y., Zhao G.-F., "A global review of deep geothermal energy exploration: from a view of rock mechanics and engineering", Geomech Geophys Geo Energy Geo Resour, 2019, 6(1):4.